\documentclass[epj]{webofc}
\usepackage[varg]{txfonts}   
\usepackage{hyperref}

\renewcommand*{\eqref}[1]{Eq.~(\ref{eq:#1})}

\wocname{ARENA-2016}
\woctitle{ARENA-2016}
\begin{document}
\title{Nanosecond-level time synchronization of AERA using a beacon reference transmitter and commercial airplanes}
%
%

\author{\firstname{Tim} \lastname{Huege}\inst{1}\fnsep\thanks{\email{tim.huege@kit.edu}}
for the \lastname{Pierre Auger Collaboration}\inst{2}}


\institute{Karlsruhe Institute of Technology (KIT), Institut f\"ur Kernphysik (IKP), Karlsruhe, Germany
\and
           Full author list: http://auger.org/archive/authors\_2016\_06.html
          }

\abstract{%
Radio detection of cosmic-ray air showers requires time synchronization of detectors on a nanosecond level, especially for advanced reconstruction algorithms based on the wavefront curvature and for interferometric analysis approaches. At the Auger Engineering Radio Array, the distributed, autonomous detector stations are time-synchronized via the Global Positioning System which, however, does not provide sufficient timing accuracy. We thus employ a dedicated beacon reference transmitter to correct for event-by-event clock drifts in our offline data analysis. In an independent cross-check of this “beacon correction” using radio pulses emitted by commercial airplanes, we have shown that the combined timing accuracy of the two methods is better than 2 nanoseconds.
}
\maketitle
\section{Introduction}

With the Auger Engineering Radio Array (AERA) \cite{SchulzIcrc2015}, we measure radio emission from cosmic-ray air showers \cite{HuegePLREP,SchroederReview} in the 30-80~MHz band. To maximize the event reconstruction potential, a highly accurate time synchronization of the individual radio detectors is required. In particular, interferometric and wavefront analyses rely on a time synchronization to within 1~ns or better. In cabled setups, nanosecond time synchronization can be achieved with established techniques. AERA, however, consists of independent, autonomous detector stations which are only connected wirelessly, making accurate time synchronization very challenging. In this article, we describe how we achieve a time synchronization of the AERA detector stations to within 2~ns using a dedicated reference transmitter, a so-called ``beacon'', together with an independent cross-check based on the measurement of radio pulses emitted by commercial airplanes. Details of the presented analysis can be found in \cite{AERAAirplane}.

\section{Time synchronization with a reference beacon}

The individual  AERA stations each use a GPS clock to synchronize the array timing. Manufacturers state timing accuracy of order 5~ns for their GPS clocks. This, however, is insufficient for interferometric and wavefront analyses. To improve the time synchronization, we thus operate a reference beacon transmitter \cite{SchroederAschBaehren2010} in the vicinity of the radio detector array. This transmitter emits monochromatic, continuous sine waves at frequencies of $58.887\,$MHz, $61.523\,$MHz, $68.555\,$MHz, and $71.191\,$MHz, i.e., within the AERA detection band of 30-80~MHz. Whenever an air-shower event is recorded with AERA, the beacon transmitter signals are thus recorded within the normal data stream. As radio pulses from air showers are broad-band, the very narrow-band signals of the beacon transmitter can be suppressed in an offline analysis \citep{AbreuAgliettaAhn2011} with minimal loss of data quality once they have been used for timing calibration.

\begin{figure*}[t]
  \centering
  \includegraphics[width=0.43\textwidth]{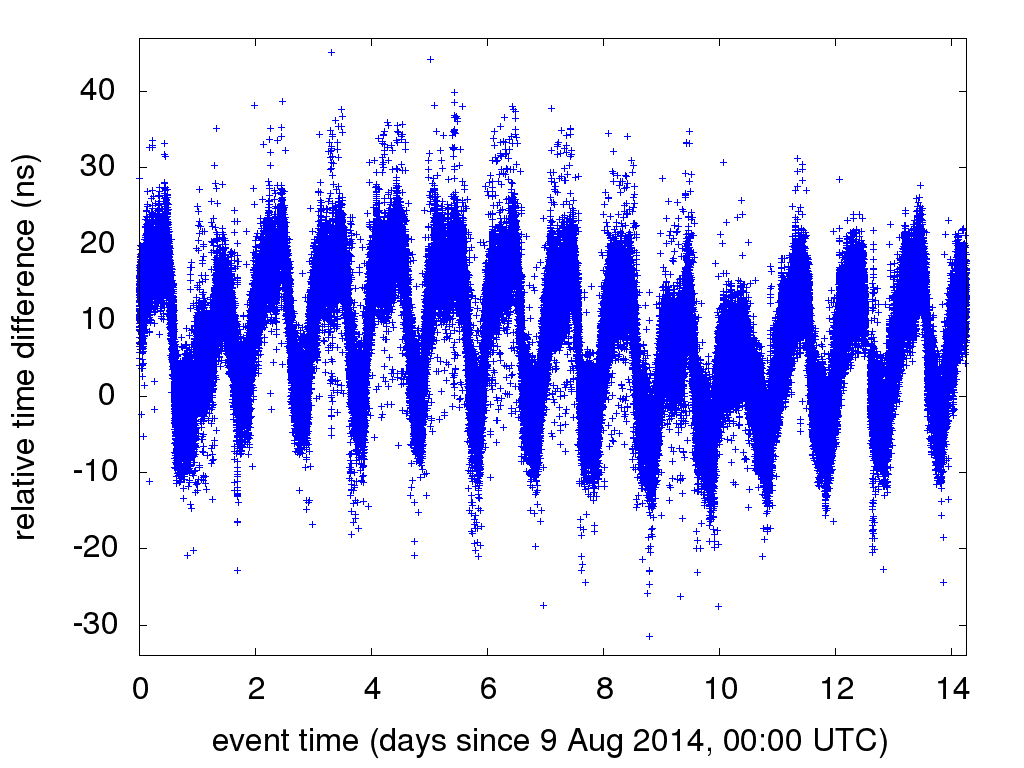}
  \includegraphics[width=0.52\textwidth]{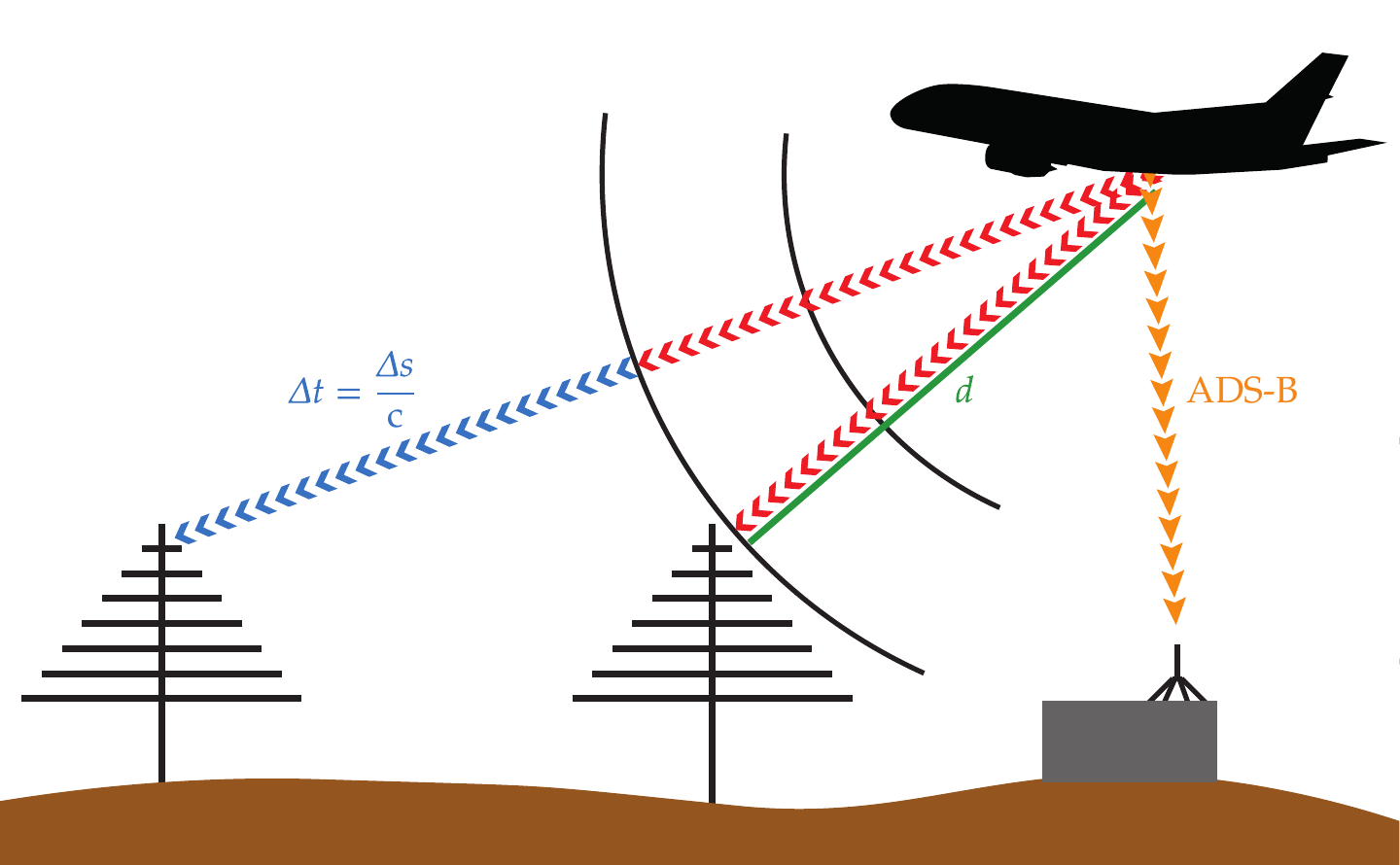}
  \caption{Left: Relative timing between two AERA stations 
  as measured using the beacon phasing. Each dot represents a measured
  offset with respect to the expectation, interpreted as a time offset between the GPS clocks of the two detector stations.
  Right: Illustration of the time calibration using commercial airplanes. The
  airplane broadcasts real-time position information via digital ADS-B packets at 
  1090~MHz, which are received and interpreted by a 
  dedicated setup in the AERA field in real time. If the airplane also 
  emits pulsed signals in the frequency 
  range of 30--80~MHz recorded by the AERA detector stations, it can be
  used as a timing calibration source.
  }
  \label{fig:GPSdrifts}
\end{figure*}

The phasing of the four frequencies expected at a given detector station with respect to the phasing at an arbitrary reference station can be calculated from the relative positions of the two stations and the position of the beacon transmitter. Any deviations from the expected phasing signify a timing offset between the two compared stations, i.e., a drift in the GPS clock of either or both of the two stations. In fact, an analysis of the measured reference phases reveals that there are very significant drifts on timescales of minutes to days, well beyond the nominal 5~ns timing accuracy of the GPS clocks, as is shown in the left panel of Fig.\ \ref{fig:GPSdrifts}.

Using the beacon phasing, the relative GPS clock drift can be measured and corrected for on a per-event level. However, it was unclear whether the beacon purely measures GPS clock drifts or can be affected by other factors, e.g., environmental conditions that could possibly influence the beacon signal transmission. An independent cross-check of this method was therefore needed.

\section{Time synchronization with airplane pulses}

To independently cross-check the beacon method, we need a source with a known position which emits pulsed radio signals in the 30-80~MHz band. These signals need to be measurable simultaneously in as many AERA stations as possible. It turns out that some commercial airplanes emit detectable radio pulses in this frequency band. The aircraft position can be determined in real-time from the Automatic Dependent Surveillance Broadcast (ADS-B) messages that modern airplanes transmit digitally at a frequency of 1090~MHz. To intercept this position information, we have installed a ``software-defined radio'' setup based on a commercial USB receiver for digital terrestrial video broadcasts (DVB-T). The total cost for the setup was below 50 USD, yet it has been running very reliably in the field for years. We use the position information for two purposes: First, we notify our data acquisition in real-time of approaching airplanes and allow repeated triggers from the corresponding direction to pass through --- normally, signals like these that are clearly not of cosmic-ray origin are filtered out by our self-trigger algorithms. Second, in an offline analysis, we use the airplane position to calculate the expected arrival times of the pulsed emissions at the individual AERA stations, which we then compare with the measured pulse arrival times. Deviations point to offsets in the timing of the individual AERA stations, in particular arising from their GPS clocks. The concept is sketched in the right panel of Fig.\ \ref{fig:GPSdrifts}.

\begin{figure*}[t]
  \centering
  \includegraphics[width=0.47\textwidth,clip=true,trim= 0cm 0cm 1.43cm 0cm]{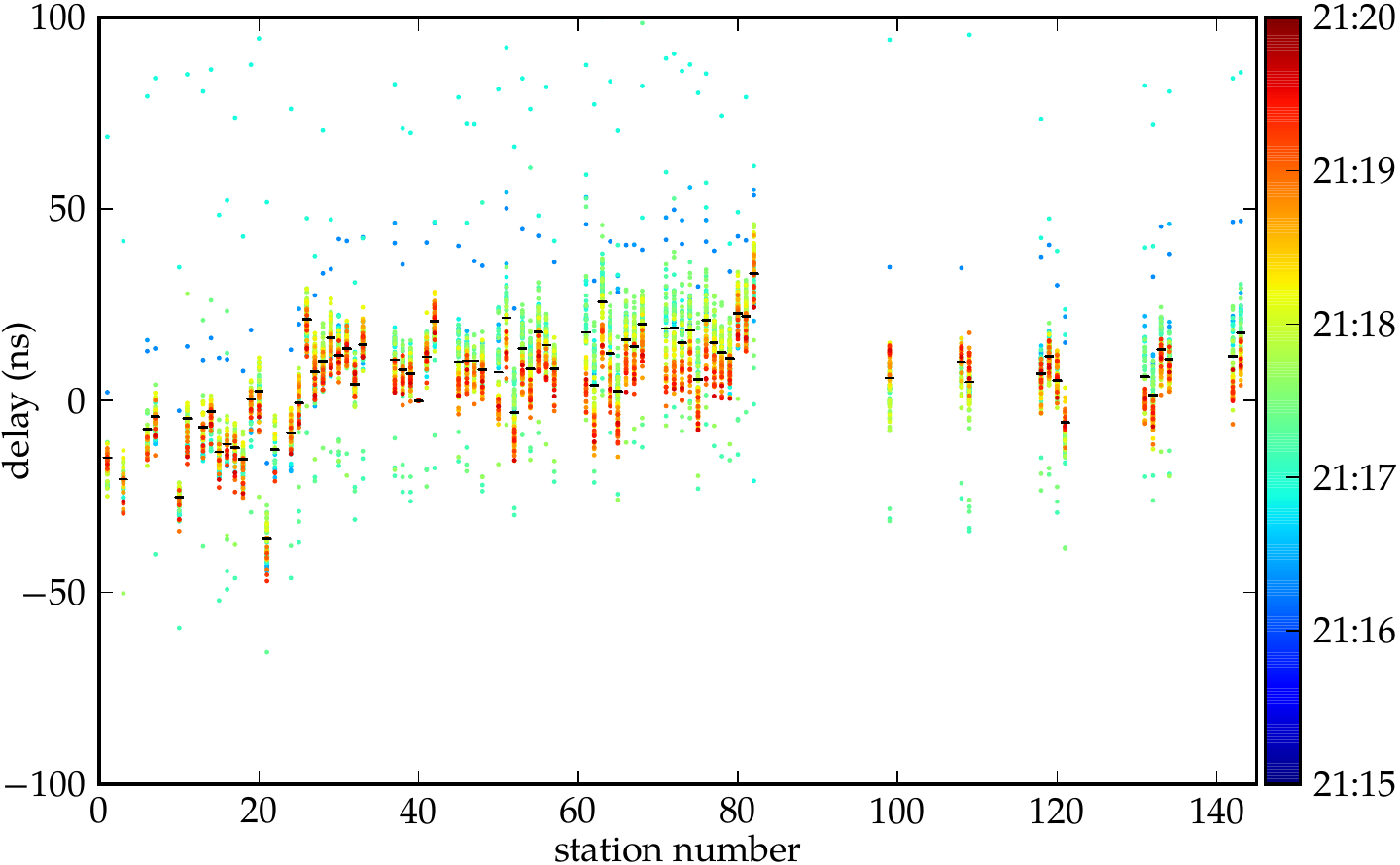}
  \includegraphics[width=0.52\textwidth]{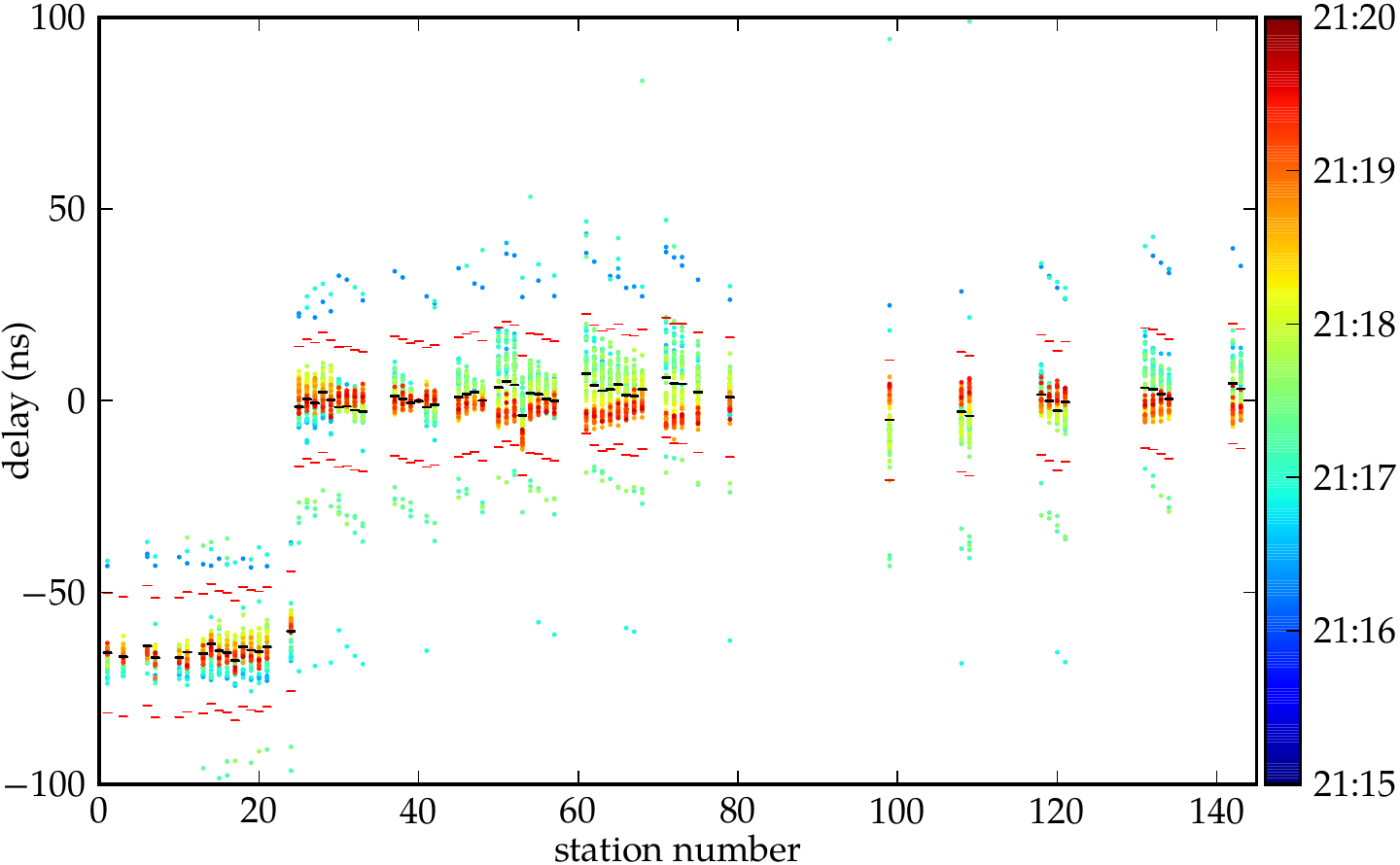}
  \caption{Left: Time offsets determined from AERA and ADS-B data for an individual airplane. Offsets are
  relative to station 40 chosen as an arbitrary reference. Gaps in the station numbering are related to 
  detector stations not participating in the airplane trigger. The color code denotes the local time
  at which the airplane pulses were detected.
  Right: Time offsets when applying the airplane timing calibration after per-event clock-drift correction using the 
  beacon method. The red lines show the range of $\mathrm{mean} {}\pm{} 4 \cdot
  \mathrm{MAD}$ (median absolute deviation), values further out are ignored for the calculation of the mean time offsets.
  }
  \label{fig:analysis}
\end{figure*}

In the left panel of Fig.\ \ref{fig:analysis}, we show the time offsets (with respect to arrival time at a chosen reference station) for each individual detector station as determined from the radio pulses emitted by a single airplane that overflew AERA over the course of approximately 5 minutes. Each individual data point marks a measurement for an airplane radio pulse emitted at the indicated local time. There is significant scatter as well as clear outliers in these data for a given station. The outliers can be explained by misidentification of the pulse maximum and thus the arrival time. The visible time-ordering of the data is not fully understood. Nevertheless, the mean of the time offsets can be determined per station. If time synchronization over the whole array were perfect, all means should be at a delay value of zero. However, the mean values scatter by tens of nanoseconds over the various stations, indicating deviations in the time synchronization of the same order as was observed in the beacon reference phasing measurements shown in the left panel of Fig.\ \ref{fig:GPSdrifts}.

In the right panel of Fig.\ \ref{fig:analysis}, the same data were analyzed. However, the beacon timing correction was applied before determining the timing offsets on the basis of the airplane analysis. If the beacon timing correction of the GPS clocks works as expected, all means should be shifted to the zero line. For stations 25 and higher, this is indeed the case. These are detector stations using Butterfly antennas \cite{AERAAntennaPaper2012}, same as the reference station chosen in this analysis (station 40). Stations 1-24 use logarithmic-periodic dipole antennas (LPDAs) \cite{AERAAntennaPaper2012} and have mean time offsets of approximately -65~ns. This means that after the beacon timing correction, indeed stations with the same type of antenna have a much better time synchronization than before the beacon correction. However, there is a previously unknown time difference between the two types of antenna, which can easily be corrected for now that it is known.

\begin{figure}
  \centering
  \sidecaption
  \includegraphics[width=0.33\textwidth,clip=true,trim= 0cm 0cm 7cm 0cm]{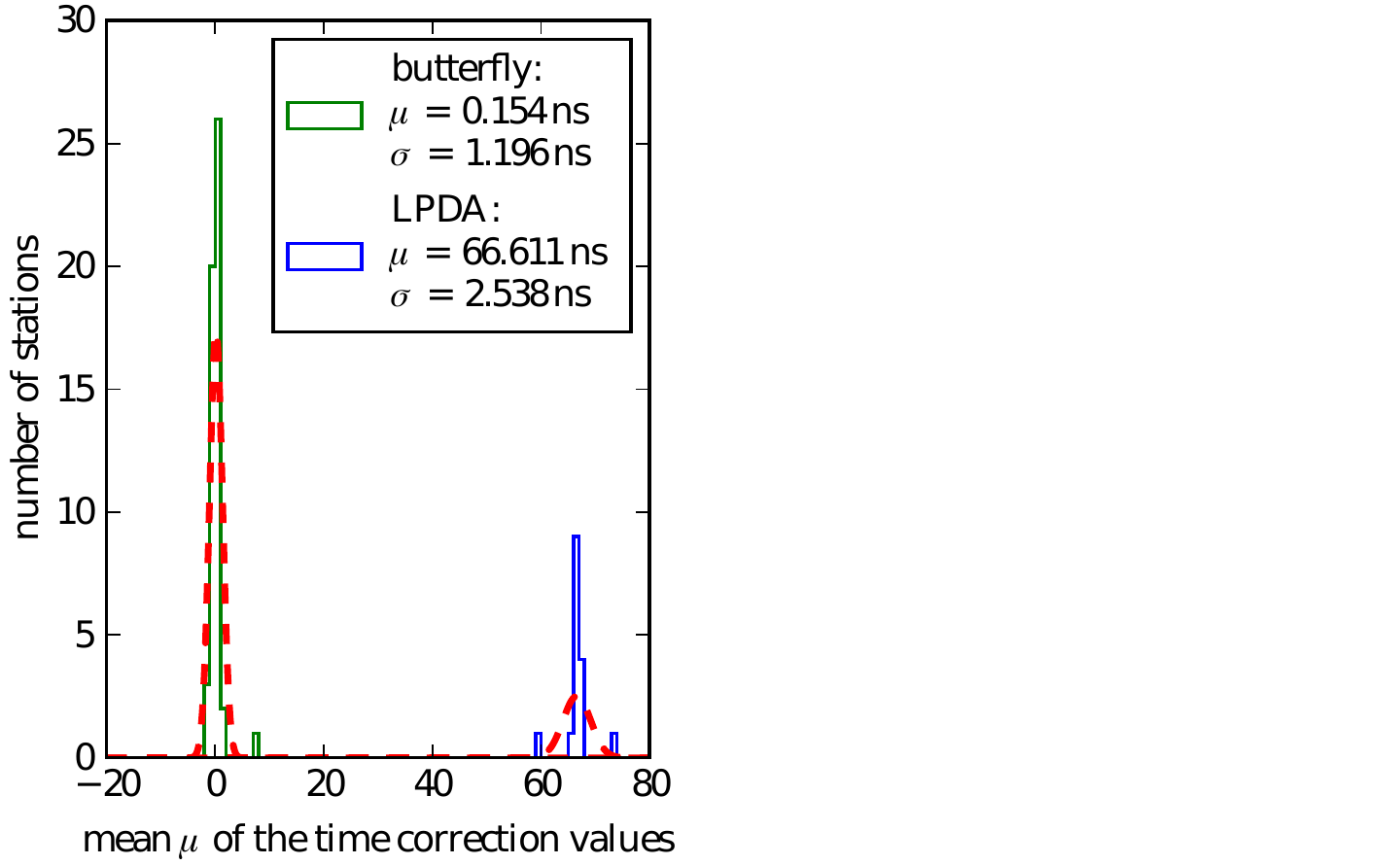}
  \caption{Histogram of the mean time correction values $\mu$ determined from the combination of beacon and airplane time synchronization methods. The standard deviations $\sigma$ of the distributions provide a measure for the average agreement between the two methods.
  The values stated in the statistics box are from fitted Gaussians (dashed lines).}
  \label{fig:histogram}
\end{figure}

A more quantitative view at the distribution of the mean values after beacon correction is given in Fig.\ \ref{fig:histogram}. The standard deviation of the mean values within the group of detector stations with the same antenna is of order 2~ns. The combined timing accuracy of both the beacon- and airplane methods is thus of order 2~ns. The airplane method is expected to have the larger systematic uncertainties (see, e.g., the time-ordering of data points), so that the beacon method alone might well be more accurate than the combined 2~ns.

\section{Conclusion}

Nanosecond-level time synchronization of distributed arrays without cable connection is a challenging task. In AERA, we have tackled this challenge with two independent approaches. A reference beacon transmitter continuously emits signals that are recorded in our normal event data stream and can be used to correct GPS clock offsets on a per-event basis. Indeed, offsets of order tens of nanoseconds are visible in our data, i.e., the nominal 5~ns accuracy of the GPS clocks is not reached in practice. To validate the beacon-correction method, we have also determined timing offsets in our detector array on the basis of radio pulses emitted by commercial airplanes, the positions of which can be determined in real-time from ADS-B transmissions. The airplane analysis confirms that the beacon timing correction works and reaches an accuracy of 2~ns or better. In addition, a systematic time offset between the LPDA and Butterfly antennas used in AERA was discovered and is now routinely corrected for.



%
%
%
%
%

\end{document}